\newcount\mgnf\newcount\tipi\newcount\tipoformule\newcount\greco 
\tipi=2          
\tipoformule=0   

\global\newcount\numsec\global\newcount\numfor
\global\newcount\numapp\global\newcount\numcap
\global\newcount\numfig\global\newcount\numpag
\global\newcount\numnf

\def\SIA #1,#2,#3 {\senondefinito{#1#2}%
\expandafter\xdef\csname #1#2\endcsname{#3}\else
\write16{???? ma #1,#2 e' gia' stato definito !!!!} \fi}

\def \FU(#1)#2{\SIA fu,#1,#2 }

\def\etichetta(#1){(\veroparagrafo.\veraformula)%
\SIA e,#1,(\veroparagrafo.\veraformula) %
\global\advance\numfor by 1%
\write15{\string\FU (#1){\equ(#1)}}%
\write16{ EQ #1 ==> \equ(#1)  }}
\def\etichettaa(#1){(A\veraappendice.\veraformula)
 \SIA e,#1,(A\veraappendice.\veraformula)
 \global\advance\numfor by 1
 \write15{\string\FU (#1){\equ(#1)}}
 \write16{ EQ #1 ==> \equ(#1) }}
\def\getichetta(#1){Fig. \verafigura
 \SIA g,#1,{\verafigura}
 \global\advance\numfig by 1
 \write15{\string\FU (#1){\graf(#1)}}
 \write16{ Fig. #1 ==> \graf(#1) }}
\def\retichetta(#1){\numpag=\pgn\SIA r,#1,{\verapagina}
 \write15{\string\FU (#1){\rif(#1)}}
 \write16{\rif(#1) ha simbolo  #1  }}
\def\etichettan(#1){(n\verocapitolo.\veranformula)
 \SIA e,#1,(n\verocapitolo.\veranformula)
 \global\advance\numnf by 1
\write16{\equ(#1) <= #1  }}

\newdimen\gwidth
\gdef\profonditastruttura{\dp\strutbox}
\def\senondefinito#1{\expandafter\ifx\csname#1\endcsname\relax}
\def\BOZZA{
\def\alato(##1){
 {\vtop to \profonditastruttura{\baselineskip
 \profonditastruttura\vss
 \rlap{\kern-\hsize\kern-1.2truecm{$\scriptstyle##1$}}}}}
\def\galato(##1){ \gwidth=\hsize \divide\gwidth by 2
 {\vtop to \profonditastruttura{\baselineskip
 \profonditastruttura\vss
 \rlap{\kern-\gwidth\kern-1.2truecm{$\scriptstyle##1$}}}}}
\def\verapagina{
{\romannumeral\number\numcap}.\number\numsec.\number\numpag}}

\def\alato(#1){}
\def\galato(#1){}
\def\veroparagrafo{\number\numsec}\def\veraformula{\number\numfor}
\def\veraappendice{\number\numapp}
\def\verapagina{\number\pageno}\def\veranformula{\number\numnf}
\def\verafigura{{\romannumeral\number\numcap}.\number\numfig}
\def\verocapitolo{\number\numcap}\def\veranformula{\number\numnf}
\def\Eqn(#1){\eqno{\etichettan(#1)\alato(#1)}}
\def\eqn(#1){\etichettan(#1)\alato(#1)}

\def\Eq(#1){\eqno{\etichetta(#1)\alato(#1)}}
\def\eq(#1){\etichetta(#1)\alato(#1)}
\def\Eqa(#1){\eqno{\etichettaa(#1)\alato(#1)}}
\def\eqa(#1){\etichettaa(#1)\alato(#1)}
\def\dgraf(#1){\getichetta(#1)\galato(#1)}
\def\drif(#1){\retichetta(#1)}

\def\eqv(#1){\senondefinito{fu#1}$\clubsuit$#1\else\csname fu#1\endcsname\fi}
\def\equ(#1){\senondefinito{e#1}\eqv(#1)\else\csname e#1\endcsname\fi}
\def\graf(#1){\senondefinito{g#1}\eqv(#1)\else\csname g#1\endcsname\fi}
\def\rif(#1){\senondefinito{r#1}\eqv(#1)\else\csname r#1\endcsname\fi}
\def\bib[#1]{[#1]\numpag=\pgn
\write13{\string[#1],\verapagina}}

\def\include#1{
\openin13=#1.aux \ifeof13 \relax \else
\input #1.aux \closein13 \fi}

\openin14=\jobname.aux \ifeof14 \relax \else
\input \jobname.aux \closein14 \fi
\openout15=\jobname.aux
\openout13=\jobname.bib


\ifnum\tipoformule=1\let\Eq=\eqno\def\eq{}\let\Eqa=\eqno\def\eqa{}
\def\equ{}\fi


{\count255=\time\divide\count255 by 60 \xdef\hourmin{\number\count255}
        \multiply\count255 by-60\advance\count255 by\time
   \xdef\hourmin{\hourmin:\ifnum\count255<10 0\fi\the\count255}}

\def\oramin{\hourmin }

\def\data{\number\day/\ifcase\month\or january \or february \or march \or
april \or may \or june \or july \or august \or september
\or october \or november \or december \fi/\number\year;\ \oramin}

\setbox200\hbox{$\scriptscriptstyle \data $}

\newcount\pgn \pgn=1
\def\foglio{\number\numsec:\number\pgn
\global\advance\pgn by 1}
\def\foglioa{A\number\numsec:\number\pgn
\global\advance\pgn by 1}

\footline={\rlap{\hbox{\copy200}}\hss\tenrm\folio\hss}

\def\TIPIO{
\font\setterm=amr7 
\def \settepunti{\def\rm{\fam0\setterm}
\textfont0=\setterm   
\normalbaselineskip=9pt\normalbaselines\rm
}\let\nota=\settepunti}

\def\TIPITOT{
\font\twelverm=cmr12
\font\twelvei=cmmi12
\font\twelvesy=cmsy10 scaled\magstep1
\font\twelveex=cmex10 scaled\magstep1
\font\twelveit=cmti12
\font\twelvett=cmtt12
\font\twelvebf=cmbx12
\font\twelvesl=cmsl12
\font\ninerm=cmr9
\font\ninesy=cmsy9
\font\eightrm=cmr8
\font\eighti=cmmi8
\font\eightsy=cmsy8
\font\eightbf=cmbx8
\font\eighttt=cmtt8
\font\eightsl=cmsl8
\font\eightit=cmti8
\font\sixrm=cmr6
\font\sixbf=cmbx6
\font\sixi=cmmi6
\font\sixsy=cmsy6
\font\twelvetruecmr=cmr10 scaled\magstep1
\font\twelvetruecmsy=cmsy10 scaled\magstep1
\font\tentruecmr=cmr10
\font\tentruecmsy=cmsy10
\font\eighttruecmr=cmr8
\font\eighttruecmsy=cmsy8
\font\seventruecmr=cmr7
\font\seventruecmsy=cmsy7
\font\sixtruecmr=cmr6
\font\sixtruecmsy=cmsy6
\font\fivetruecmr=cmr5
\font\fivetruecmsy=cmsy5
\textfont\truecmr=\tentruecmr
\scriptfont\truecmr=\seventruecmr
\scriptscriptfont\truecmr=\fivetruecmr
\textfont\truecmsy=\tentruecmsy
\scriptfont\truecmsy=\seventruecmsy
\scriptscriptfont\truecmr=\fivetruecmr
\scriptscriptfont\truecmsy=\fivetruecmsy
\def \eightpoint{\def\rm{\fam0\eightrm}
\textfont0=\eightrm \scriptfont0=\sixrm \scriptscriptfont0=\fiverm
\textfont1=\eighti \scriptfont1=\sixi   \scriptscriptfont1=\fivei
\textfont2=\eightsy \scriptfont2=\sixsy   \scriptscriptfont2=\fivesy
\textfont3=\tenex \scriptfont3=\tenex   \scriptscriptfont3=\tenex
\textfont\itfam=\eightit  \def\it{\fam\itfam\eightit}%
\textfont\slfam=\eightsl  \def\sl{\fam\slfam\eightsl}%
\textfont\ttfam=\eighttt  \def\tt{\fam\ttfam\eighttt}%
\textfont\bffam=\eightbf  \scriptfont\bffam=\sixbf
\scriptscriptfont\bffam=\fivebf  \def\bf{\fam\bffam\eightbf}%
\tt \ttglue=.5em plus.25em minus.15em
\setbox\strutbox=\hbox{\vrule height7pt depth2pt width0pt}%
\normalbaselineskip=9pt
\let\sc=\sixrm  \let\big=\eightbig  \normalbaselines\rm
\textfont\truecmr=\eighttruecmr
\scriptfont\truecmr=\sixtruecmr
\scriptscriptfont\truecmr=\fivetruecmr
\textfont\truecmsy=\eighttruecmsy
\scriptfont\truecmsy=\sixtruecmsy
}\let\nota=\eightpoint}

\newfam\msbfam   
\newfam\truecmr  
\newfam\truecmsy 
\newskip\ttglue
\ifnum\tipi=0\TIPIO \else\ifnum\tipi=1 \TIPI\else \TIPITOT\fi\fi

\global\newcount\numpunt

\magnification=\magstephalf
\baselineskip=16pt
\parskip=8pt

\def\a{\alpha}

\def\d{\delta}
\def\e{\epsilon}
\def\f{\phi}

\def\s{\sigma}

\def\z{\zeta}

\def\E{{I\kern-.25em{E}}}
\def\N{{I\kern-.22em{N}}}
\def\M{{I\kern-.22em{M}}}
\def\R{{I\kern-.22em{R}}}
\def\Z{{Z\kern-.5em{Z}}}
\def\1{{1\kern-.25em\hbox{\rm I}}}
\def\eu{{1\kern-.25em\hbox{\sm I}}}
\def\f1{{1\kern-.25em\hbox{\vsm I}}}
\def\C{{C\kern-.75em{C}}}
\def\P{{I\kern-.25em{P}}}





\def\chap #1#2{\line{\ch #1\hfill}\numsec=#2\numfor=1}


\newcount\foot
\foot=1
\def\note#1{\footnote{${}^{\number\foot}$}{\ftn #1}\advance\foot by 1}

\def\frac#1#2{{#1\over #2}}

\def\text#1{\quad{\hbox{#1}}\quad}
\def\newpage{\vfill\eject}

\def\thanks{\noindent{\bf Acknowledgements: }}



\font\ch=cmbx12
\font\ftn=cmr8

\font\it=cmti10
\font\bf=cmbx10
\font\sm=cmr7
\font\vsm=cmr6


\font\tit=cmbx12
\font\aut=cmbx12
\font\aff=cmsl12
\def\s{\char'31}
\nopagenumbers
{$  $}
\vskip1.5truecm
\centerline{\tit COMMENT ON }
\vskip.2truecm
\centerline{\tit ``CAPACITY OF THE HOPFIELD MODEL''\footnote{${}^\#$}{\ftn 
Work
partially supported by the Commission of the European Communities
under contract  CHRX-CT93-0411}}
\vskip1.5truecm
\centerline{\aut Anton Bovier 
\footnote{${}^1$}{\ftn e-mail:
bovier@wias-berlin.de}
}
\vskip.1truecm
\centerline{\aff Weierstra\s {}--Institut}
\centerline{\aff f\"ur Angewandte Analysis und Stochastik}
\centerline{\aff Mohrenstrasse 39, D-10117 Berlin, Germany}
\vskip.4truecm
\vskip4truecm\rm
\def\s{\sigma}
\noindent {\bf Abstract:} In a recent paper ``The capacity of the Hopfield 
model, J. Feng and B. Tirozzi claim to prove rigorous results on 
the storage capacity that are in conflict with the predictions of 
the replica approach. We show that their results are in error and that their
approach, even when the worst mistakes are corrected, is not giving any 
mathematically rigorous results.


$ {} $

\newpage
\count0=1
\footline{\hss\folio\hss}

\chap{}1

The paper [FT] by Feng and Tirrozi addresses the interesting question of the 
storage capacity of the Hopfield model. This value, namely the maximal ratio 
between the number of patterns to the number of neurons for which
the Hopfield model works as a memory has been first observed numerically by 
Hopfield to be about $0.14$, and a value close to that was obtained 
analytically
by Amit et al [AGS] with the use of the replica trick. Refined estimates 
using replica symmetry breaking schemes were obtained later. However,
the replica trick is mathematically not rigorous, and there have been many
attempts to obtain 
such results in a mathematically rigorous way. The best results in this 
respect so far were rigorous lower bounds on $\a_c$ by Newman [N]  
which recently have been improved by Loukianova [L1] and Talagrand [Ta].  
These bounds are still by at least 50 per cent off the expected value.
Obtaining upper bounds on $\a_c$ has proven, a much more difficult issue,
and the only result to our knowledge was obtained very recently by
Loukianova [L2], who proved that for any $\a>0$, the minimum of the 
Hamiltonian is at some finite distance away from the patterns, and 
that as $\a$ tends to infinity, this distance tends to at least $0.05$.
Although her proof is very nice and interesting, the numerical values 
are of course far from satisfactory. 

The main result of 
Feng and Tirozzi claim to prove ``rigorously'' is that, if a fraction $\d $
of errors in the retrieval is allowed, then the critical $\a=\a(\d)$ is 
given by $\a(\d)=\frac {(1-2\d)^2}{(1-\d)^2}$. 

This result appears obviously false,
since it gives $\a(\d)$ close to one if $\d$ is chosen 
close to $0$, and $\a(\d)$ close to zero, if
$\d$ is close to $1/2$, contrary to what {\it has} to be the case.
One might be first tempted to believe that this formula 
for $\a(\d)$ is a misprint, but it is repeated consistently in the paper,
and moreover, based on this formula, the authors argue that ``the replica 
trick approach to the capacity of the Hopfield model is only valid in the case
$\a_N\rightarrow 0(N\rightarrow \infty)$''.

Given the very strong and surprising claims made in this paper, it 
appears worthwhile to analyze their ``rigorous proof'' in some detail 
in order to avoid misconceptions.

Feng and Tiriozzi study the fixpoints of a deterministic gradient 
dynamic of the Hopfield model, i.e. solutions of the 
equation 
$$
\s_i=\hbox{sign}\left(\frac 1N\sum_{\mu=1}^p\sum_{j\neq i}\xi_i^\mu\xi_j^\mu\s_j\right)
\Eq(1)
$$
Since they are interested in solutions ``near'' one, say the first, pattern, 
$\xi^1$, it is reasonable to index the configurations $\s$ by the set 
$B\subset \{1,\dots,N\}$
on which they differ from  $\xi^1$, i.e. set\note{I regret to 
have to introduce some 
notation that is different from that of Feng and Tirozzi.}
$$
\s^B_i\equiv\cases{\xi^1_i,&if $i\not\in B$\cr
                   -\xi_i^1,&if $i\in B$.}
\Eq(2)
$$
The fixpoint equation \eqv(1)
can then be written in the form
$$
\s_i^B\left(\frac 1N\sum_{\mu=1}^p\sum_{j\neq i}\xi_i^\mu\xi_j^\mu\s^B_j\right)
\geq 0\text{for all $1\leq i\leq N$}
\Eq(3)
$$
After some elementary algebra, we can rewrite this as
$$
\s_i^B\left(\frac 1N\sum_{\mu=2}^p\sum_{j\neq i}\xi_i^\mu\xi_j^\mu\xi^1_j
-\frac 2N \sum_{\mu=2}^p\sum_{j\in B,j\neq  i}\xi_i^\mu\xi_j^\mu\xi^1_j
+\xi^1_i(1-|B|/N)\right)
\Eq(4)
$$
where $|B|$ denotes the cardinality of the set $B$.
Let us define the random variables\note{In [FT] these are defined in (4), 
and given 
the strange name $\scriptstyle g (N,p(N))$, which makes reference 
neither to their 
dependence on
the index $\scriptstyle i$ nor the set $\scriptstyle B$. 
We need these attributes to make meaningful 
statements later.}
$$
Z_i(N,B)\equiv\xi_i^1\left(\frac 1N \sum_{\mu=2}^{p(N)}\sum_{j=1,j\neq i}
\xi_i^\mu\xi_j^\mu\xi_j^1-2
\frac 1N \sum_{\mu=2}^{p(N)}\sum_{j\in B,j\neq i}
\xi_i^\mu\xi_j^\mu\xi_j^1\right)
\Eq(5)
$$
Then \eqv(4) can be written as 
$$
\eqalign{
Z_i(N,B)&\geq -(1-2|B|/N),\text {if $i\not\in B$}\cr
Z_i(N,B)&\leq -(1-2|B|/N),\text {if $i\in B$}
}\Eq(6)
$$
The arguments of [FT] relies basically on their observation in Eq. (4) that
for any fixed $B$, the random variables $Z_i(N,B)$ 
converge in distribution to $-\sqrt \a \z_i$\note{The minus sign appearing here
is rather unconventional, given that $\scriptstyle \z_i$ and 
$\scriptstyle-\z_i$ have the 
same distribution. However, it plays a r\^ole in the course of the mistakes 
they make later.} 
(assuming that $p(N)/N$ tends to $\a$),
``by the central limit theorem'', where the $\z_i$ are i.i.d.
standard normal variables. 
The remainder of their analysis is then based on the
study of the distribution of the $k$-th maxima of i.i.d. gaussian 
random variables.

This procedure involves an interchange of limits that is not justified.
From the CLT, one obtains the convergence of the 
variables in \eqv(1) in the sense that for a given $B$,
{\it for any fixed, finite set $I$ of 
indices}, the family $\{Z_i(N,B)\}_{i\in I}$ 
converges in distribution to a family of i.i.d. gaussians.
This does {\it not} imply that e.g. $\max_{i=1}^N Z_i(N,B)$
converges, e.g. in distribution to the same limit as 
$\max_{i=1}^N \sqrt\a \z_i$!
Maxima are not continuous functions w.r.t. the product topology, and therefore 
 convergence in distribution and taking of maxima cannot be interchanged.
(Take the following example: Let $X_i(N)\equiv \z_i$, if $i<N$, and 
$X_N(N)=N$. This family converges to i.i.d. standard gaussians, as 
$N$ tends to infinity, but the maximum
(which is always $N$) is totally different from the maximum of $N$ i.i.d.
gaussians, which behaves like $\sqrt {\ln N}$!). 

One should keep in mind that it is precisely this difficulty that has 
prevented reasonable upper bounds on $\a_c$ in the past. 
Loukianova, for instance uses a very clever idea of ``negative association''
to compare the {\it dependent } variables $Z_i(N,B)$ to independent ones,
but for this a price had to be paid that prevented sharp estimates.

While at this point it is clear that the arguments in [FT] are in no way
``rigorous'', it may be still interesting to follow the sequel of their 
arguments in some more detail. Let us first consider the case of 
``perfect retrieval'', Subsection 3.1. Here $B$ is the empty set and
their result relies on the assumption that the distribution of the 
maximum over $i$ of the $Z_i(N,\emptyset)$ converges to that 
of i.i.d. gaussians, which is not justified. There is no surprise in the
fact that they obtain the same result as MacEliece et al. [MPRV], because the
heuristic arguments of [MPRV] (which do not claim to give a rigorous proof!)
are  identical to those put forward here.  I should stress that
to my knowledge there is no rigorous proof of the ``if and only if'' statement.

In subsection 3.2 [FT] study the case of non-perfect retrieval, i.e.
they look for the critical 
$\a(\d)$ such that a fixpoint $\s^B$ will exist with $|B|=\d N$. The way they 
 seem to argue is as follows: $Z_i(N,B)/\sqrt\a$ converges to 
family of i.i.d. gaussian r.v.'s. When what is the 
fraction of $N$ i.i.d. gaussians that is larger than $x$? If this number is
$g(x)$, then $g(-(1-2\d)/\sqrt\a)$ of the $Z_i(N,B)$ will be larger 
than $(1-2\d)$, and so we will find a set $B$ of size $\d N$ precisely when
$(1-\d) = g(-(1-2\d)/\sqrt\a)$! 
At this point  the authors claim
(see the first phrase on  page 3386, and figure 1) that the 
$[xN]$-largest of the $N$ gaussians $\z_i$ is of order $x$, so that 
$[x N]$ of them would be smaller than $ x$, 
i.e. $[xN]$ of the $-\sqrt\a \z_i$
would be larger than $-\sqrt \a x$ which means that they take $g(x)=-x$. 
It does not become clear where they draw this claim
but it is clear that it is totally wrong and leads to  their absurd 
main result.

One may ask whether their arguments  can be improved.
First, how can we compute the size of the set of indices for which 
$Z_i(N,B)\geq x$? Obviously, one would want to study the quantity
$$
G_N(x,B)\equiv\frac 1N\sum_{i=1}^N \1_{Z_i(N,B)\geq x}
\Eq(7)
$$
which is nothing but (one minus) the distribution function of the empirical measure
of the variables $Z_i(N,B)$. Now, if we replaced all the
$Z_i(N,B)$ by $\sqrt\a \z_i$ (I take the freedom to drop their pointless
minus sign), then by the strong law of large numbers
$$
\lim_{N\uparrow\infty}\frac 1N\sum_{i=1}^N \1_{\sqrt a\z_i\geq x}
=\frac 1{\sqrt{2\pi}}\int_{x/\sqrt\a}^\infty e^{-\frac {y^2}{2}}dy
\equiv\Phi(-x/\sqrt\a), \text{a.s.}
\Eq(8)
$$
This would then yield the somewhat more reasonable looking result 
$\a(\d)=\frac {(1-2\d)^2}{(\Phi^{-1}(1-\d))^2}$. But is that result to be 
trusted?  First, the CLT again cannot justify the passage 
to the gaussians, because just as the maxima, the 
empirical measure is not a continuous function.
If one is somewhat optimistic, one may hope  to prove convergence of
$G_N(x,B)$ to a gaussian distribution function, {\it for fixed $B$}.
(E.g. Talagrand [Ta]  proves this for the case $B=\emptyset$). But even that
would by no means  
imply  that $Z_i(N,B)$ itself is getting independent of 
$B$, as $N$ increases, and that the set $B'(B)$ for which 
$Z_i(N,B)\geq -(1-2\d )$ would coincide with $B$. One might want to argue 
that there should be a set $B$ for which $B'(B)=B$, but then there is no 
reason why for this {\it random} set the convergence of the empirical 
measure should hold. 
 To make such a 
statement, one would at least have to get control on the convergence of 
$G_N(x,B)$ {\it uniformly } in the different possible sets $B$, 
i.e. we should need some estimate
like
$$
\P\left[ \sup_{B\subset \{1,\dots,N\}} \|G_N(x,B)
-\Phi(x)\|>\e\right] \downarrow 0
\Eq(9)
$$
for all $\e>0$, for some norm $\|\cdot\|$. It does not seem likely that 
such a result is true, let alone that it 
can be proven. 
Note that the main difficulty here is that the number of sets $B$ is 
exponentially large, and precisely this fact is responsible for 
the relatively poor lower bounds on $\a_c$ that exist in the literature.

In conclusion,  the paper by Feng and Tirozzi has unfortunately not 
contributed to progress in the mathematical understanding of this 
interesting and challenging problem. Even if the most obvious mistakes
are corrected there remain fundamental problems in the basic approach, 
and even the improved prediction on $\a(\d)$ is no more rigorous and 
rather less 
convincing 
  than the predictions of the replica approach. 

\bigskip

\chap{References}4
\item{[AGS]} D.J. Amit, H. Gutfreund and H.
Sompolinsky, ``Statistical mechanics of neural networks near saturation'',
Ann. Phys. {\bf 173}, 30-67 (1987).
\item{[FT]} J. Feng and B. Tirozzi, ``Capacity of the Hopfield model'', 
J. Phys. A {\bf 30}, 3383-3391 (1997).
\item{[N]}  Ch.M. Newman, ``Memory capacity in neural network models:
Rigorous results'', Neural Networks {\bf 1}, 223-238 (1988).   
\item{[L1]} D. Loukianova, ``Capacit\'e de m\'emoire dans le mod\`ele 
de Hopfield'', C.R.A.S. Paris {\bf t. 318, S\'erie 1}, 157-160 (1994). 
\item{[L2]} D. Loukianova, ``Lower bounds on the restitution error in the 
Hopfield model'', Prob. Theor. Rel. Fields {\bf 107}, 161-176 (1997)
\item{[Ta]} M. Talagrand, ``Rigorous results on the Hopfield model with 
many patterns'', to appear in Prob. Theor. Rel. Fields. (1997). 

\end